\documentclass[sigconf]{acmart}
\renewcommand\footnotetextcopyrightpermission[1]{} 
\usepackage{cleveref}
\usepackage{bm}
\usepackage{bbm}
\usepackage{caption}
\usepackage{subcaption}
\usepackage{booktabs,amsfonts,dcolumn}
\usepackage{mathtools}
\usepackage{wasysym}
\usepackage{hhline}
\usepackage{float}
\usepackage{commath}

 \usepackage{multirow}
 \usepackage[ruled,vlined,linesnumbered]{algorithm2e}
\usepackage{booktabs,siunitx}

\usepackage{stmaryrd}
\usepackage{trimclip}
\usepackage{pifont}
\usepackage{enumitem}
\usepackage{mleftright}
\usepackage{scalerel,stackengine}
\usepackage{bbold}
\usepackage{placeins}
\usepackage{tcolorbox}
\tcbuselibrary{listings}
\setcopyright{acmlicensed}
\copyrightyear{2018}
\acmYear{2018}
\acmDOI{XXXXXXX.XXXXXXX}
\acmConference[Conference acronym 'XX]{Make sure to enter the correct
  conference title from your rights confirmation email}{June 03--05,
  2018}{Woodstock, NY}
\acmISBN{978-1-4503-XXXX-X/18/06}

\begin{document}
\title{Generalized Pseudo-Relevance Feedback}

\author{Yiteng Tu}
\affiliation{%
    \institution{DCST, Tsinghua University}
    \city{Beijing}
    \country{China}
}
\email{tyt24@mails.tsinghua.edu.cn}

\author{Weihang Su}
\affiliation{%
    \institution{DCST, Tsinghua University}
    \city{Beijing}
    \country{China}
}

\author{Yujia Zhou}
\affiliation{%
    \institution{DCST, Tsinghua University}
    \city{Beijing}
    \country{China}
}

\author{Yiqun Liu}
\affiliation{%
    \institution{DCST, Tsinghua University}
    \city{Beijing}
    \country{China}
}

\author{Fen	Lin}
\affiliation{%
    \institution{Tencent}
    \city{Beijing}
    \country{China}
}

\author{Qin	Liu}
\affiliation{%
    \institution{Tencent}
    \city{Beijing}
    \country{China}
}

\author{Qingyao Ai\footnotemark}
\affiliation{%
    \institution{DCST, Tsinghua University}
    \city{Beijing}
    \country{China}
}
\email{aiqy@tsinghua.edu.cn}

\begin{abstract}
Query rewriting is a fundamental technique in information retrieval (IR).
It typically employs the retrieval result as relevance feedback to refine the query and thereby addresses the vocabulary mismatch between user queries and relevant documents. 
Traditional pseudo-relevance feedback (PRF) and its vector-based extension (VPRF) improve retrieval performance by leveraging top-retrieved documents as relevance feedback. 
However, they are constructed based on two major hypotheses: the relevance assumption (top documents are relevant) and the model assumption (rewriting methods need to be designed specifically for particular model architectures). 
While recent large language models (LLMs)-based generative relevance feedback (GRF) enables model-free query reformulation, it either suffers from severe LLM hallucination or, again, relies on the relevance assumption to guarantee the effectiveness of rewriting quality. 
To overcome these limitations, we introduce an assumption-relaxed framework: \textit{Generalized Pseudo Relevance Feedback} (GPRF), which performs model-free, natural language rewriting based on retrieved documents, not only eliminating the model assumption but also reducing dependence on the relevance assumption.
Specifically, we design a utility-oriented training pipeline with reinforcement learning to ensure robustness against noisy feedback. %and relax the relevance assumption. 
Extensive experiments across multiple benchmarks and retrievers demonstrate that GPRF consistently outperforms strong baselines, establishing it as an effective and generalizable framework for query rewriting.
\end{abstract}


\maketitle

\section{Introduction}
Search engines have become indispensable tools for accessing information, powering applications ranging from web search and e-commerce to open-domain question answering and knowledge-grounded dialogue~\cite{karpukhin2020dense,ai2017learning,zhao2020knowledge,su2024wikiformer}. 
A central goal of these systems is to bridge the gap between user queries and vast document collections~\cite{li2025query,zhu2023large,shen2023large}.
However, a long-standing challenge in search engines lies in the vocabulary mismatch problem: users often express information needs with general and ambiguous terms, while relevant documents may employ more formal, specialized, or emergent terminology~\cite{li2025query,zhu2023large}. 
To address this gap, query rewriting has emerged as a crucial technique, enriching initial queries with semantically related or contextually grounded expressions to enhance the likelihood of retrieving documents that align with the user's intent~\cite{rocchio1971relevance,abdul2004umass}. 
Over decades of research, it has proven to be an effective method for improving retrieval effectiveness in both classical and neural search paradigms~\cite{gao2023precise}.

Query rewriting approaches typically first perform an initial retrieval using the original query and then refine it with top-ranked retrieved results, regarding them as the relevance feedback.
One of the most well-known paradigms is pseudo-relevance feedback (PRF)~\cite{lavrenko2017relevance,rocchio1971relevance,voorhees1994query,abdul2004umass}.
%A representative approach, RM3~\cite{abdul2004umass}, 
It estimates term distributions from the top retrieved documents, assuming that they are relevant, and interpolates them with the original query term distribution, thereby improving retrieval performance and robustness of sparse retrievers like BM25~\cite{robertson2009probabilistic}.
Vector-based pseudo-relevance feedback (VPRF)~\cite{li2023pseudo,li2025llm}, on the other hand, is a variant of PRF tailored for dense retrieval~\cite{karpukhin2020dense,zhan2020repbert,xiong2020approximate,izacard2021unsupervised} scenarios, which directly aggregates dense embeddings of the top-retrieved documents to refine the query representation. 
%For dense retrieval that maps queries and documents into a shared embedding space using neural  encoders~\cite{karpukhin2020dense,zhan2020repbert,xiong2020approximate,izacard2021unsupervised}, researchers have revisited PRF in the dense semantic space. 
%The vector-based PRF (VPRF)~\cite{li2023pseudo,li2025llm} directly aggregates dense embeddings of the top-retrieved documents to refine the query representation. 
It effectively leverages the semantic richness of neural representations and has been shown to boost retrieval effectiveness across a range of tasks~\cite{li2025llm}.

Despite their effectiveness, both PRF and VPRF are fundamentally constrained by two strong assumptions that severely limit their robustness and generalizability. 
The first one is \textbf{relevance assumption}, which assumes that all the top-ranked documents retrieved in the initial stage should be relevant and thus be beneficial for query rewriting.
While this assumption may hold in carefully curated test collections, it is far from true in real-world scenarios, where retrieval systems are inherently imperfect and top results often include noisy and irrelevant information~\cite{tu2025robust,wang2024astute}. 
Once these noisy and off-topic documents are incorporated into the rewriting process, they can introduce misleading content and even drift the reformulation away from the user’s true intent.
The second assumption is \textbf{model assumption}: these methods are tightly coupled to a specific retriever's internal representations. 
%in these methods, query rewriting must be conducted in certain retriever’s corresponding representation space.
%That is, the rewritten query is expected to be interpretable and effective only within the lexical or semantic space of the retriever that provides the feedback. 
By operating at the level of term weights or dense embeddings, the rewritten query is inherently tied to a particular model's feature space, making it challenging to transfer across different or evolving retrieval systems. 
This rigidity and coupling not only narrow their applicability but also constrain the exploration of richer, more flexible reformulation strategies~\cite{zhu2023large}.
The reliance on these two assumptions makes PRF and VPRF highly vulnerable to noisy feedback and difficult to adapt across models, motivating the search for alternative approaches. % that can relax both assumptions by operating directly at the natural language level, where user intent can be expressed and preserved more faithfully.

Recently, the rise of large language models (LLMs)~\cite{achiam2023gpt,dubey2024llama,bai2023qwen} has led to a new type of methods named generative relevance feedback (GRF)~\cite{gao2023precise,shen2023large,mackie2023generative,wang2023query2doc}.
Given a short or ambiguous query, an LLM can synthesize pseudo-documents or detailed answer-style passages that articulate the user’s information need with richer context~\cite{li2025query,gao2023precise,wang2023query2doc}.
By operating at the natural language level instead of adjusting weights or embeddings, GRF mitigates the model assumption: the reformulated query is no longer bound to a specific embedding space, making the approach more interpretable and transferable across different retrieval models and domains.
However, GRF methods still rest on the relevance assumption, assuming that the generated expansions faithfully reflect the user’s intent and provide useful retrieval cues. 
In practice, this assumption is also questionable, as LLMs are prone to hallucination, producing fluent but factually incorrect or semantically irrelevant content~\cite{su2024mitigating,su2024unsupervised}.
%Consequently, while GRF represents an important step toward more flexible and semantically grounded query reformulation, its dependence on the fragile relevance assumption continues to pose a significant barrier to the reliable deployment of such systems.

\begin{figure*}[t]
    \centering
    \includegraphics[width=0.98\textwidth]{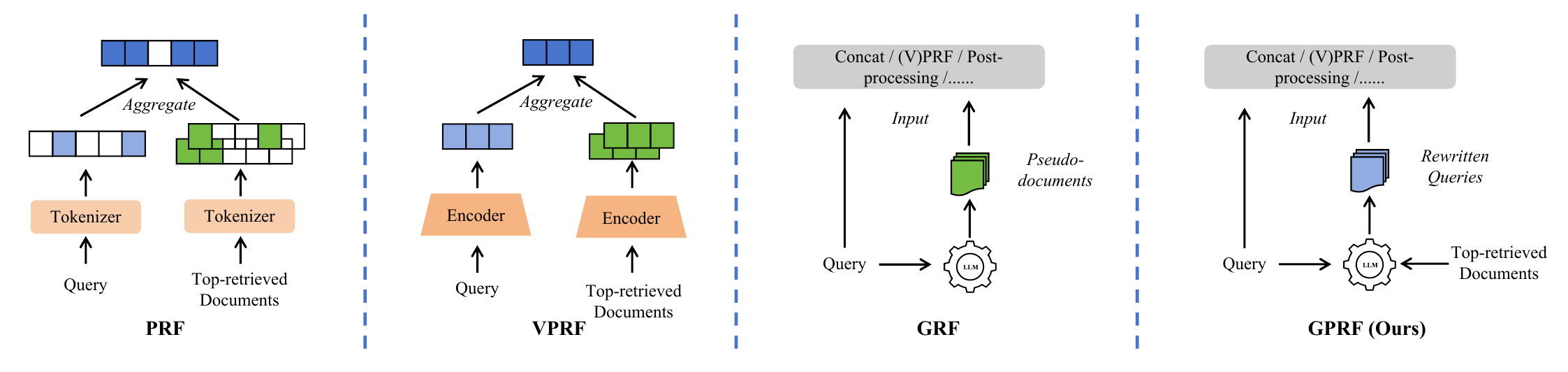}
    \caption{The comparison between PRF, VPRF, GRF, and our proposed GPRF. The pseudo-documents or rewritten queries produced by GRF and GPRF can be processed in various ways, such as directly concatenating them with the original query, integrating them into PRF or VPRF systems, or performing retrieval separately for each, then post-processing the results.}
    \label{fig:frame}
\end{figure*}

To address the limitations above, we introduce \textit{Generalized Pseudo-Relevance Feedback} (GPRF), a generative, evidence-guided query rewriting framework that relaxes both assumptions. %an \textbf{assumption-free} query rewriting framework that unifies PRF and GRF.
From the model perspective, GPRF overcomes the limitation of PRF by leveraging LLMs to conduct natural language-based query reformulation and avoids the hallucination problem of GRF by grounding the process with top-retrieved documents (as shown in Figure~\ref{fig:frame}).
From the relevance perspective, GPRF relaxes the assumption on top-retrieved documents' quality by introducing a comprehensive, utility-oriented optimization pipeline.
This pipeline is specifically designed to make the generative model robust to noisy feedback through three stages: retrieval-augmented rejection sampling filters unfaithful generations and selects high-quality training samples, supervised fine-tuning equips the model with the initial ability to generate high-quality rewrites, and reinforcement learning directly aligns the model with retrieval utility. 
By explicitly forcing the model to learn from reliable feedback in the earlier stage and then shaping its generation behavior with task-aligned rewards, our training pipeline corrects the model’s tendency to propagate misleading and detrimental information.
This process empowers the model to discern and leverage useful signals even from imperfect feedback, thus substantially mitigating the negative impact of irrelevant documents.
Extensive experiments across multiple retrievers and benchmarks demonstrate that GPRF consistently outperforms strong baselines, including both classical PRF methods and recent GRF approaches. 
These results highlight the effectiveness of our method, positioning GPRF as a promising direction for advancing query reformulation in retrieval systems.

In summary, this paper makes three key contributions:
(1) We conduct a systematic analysis of existing query rewriting methods, including PRF and GRF, highlighting two major challenges they face: the reliance on \textbf{relevance assumption} and the \textbf{model assumption}.
(2) We propose Generalized Pseudo-Relevance Feedback (GPRF) and a corresponding utility-oriented training pipeline, which effectively integrates the advantages of PRF and GRF while alleviating their weaknesses.
(3) Extensive experiments show that our framework consistently outperforms strong baselines, demonstrating its effectiveness and generalizability.

\section{Preliminary}
\subsection{Sparse Retrieval and Dense Retrieval}
We consider the standard ad-hoc retrieval setting, where a system takes as input a query $q$ and ranks documents from a large collection $\mathcal{D} = \{d_1, d_2, \dots, d_N\}$. 
The retrieval process relies on a scoring function $s(q, d)$ that estimates the relevance between the query and a document. 
%The ranked list is then obtained by sorting documents according to their scores.
Traditional sparse retrieval methods, such as BM25, represent queries and documents as high-dimensional sparse vectors over the vocabulary space $\mathcal{V}$.
Each dimension corresponds to a term, weighted by functions such as term frequency (TF) and term frequency–inverse document frequency (TF-IDF). 
The relevance score is computed by lexical matching:
\begin{equation}
s_{\text{sparse}}(q, d) = \sum_{t \in q \cap d} w_{q}(t) \cdot w_{d}(t),
\end{equation}
where $w_{q}(t)$ and $w_{d}(t)$ denote the weights of term $t$ in the query and document representations, respectively. 
%In BM25, these weights incorporate document length normalization and inverse document frequency (IDF) statistics. 
While effective and interpretable, sparse retrieval is inherently limited to surface-level term overlap and often fails to capture semantic similarity.

In contrast, dense retrieval encodes queries and documents into low-dimensional dense vectors using a neural encoder $\text{Enc}(\cdot)$. %, often derived from PLMs or LLMs. 
Each query and document is mapped into the same semantic space (denoted as $\mathbf{q}$ and $\mathbf{d}$), and their relevance is estimated via similarity measures such as inner product or cosine similarity:
\begin{equation}
    \mathbf{q} = \text{Enc}(q), \quad \mathbf{d} = \text{Enc}(d),
\end{equation}
\begin{equation}
    s_{\text{dense}}(q, d) =  \langle \mathbf{q}, \mathbf{d} \rangle,
\end{equation}
where $\langle \cdot, \cdot \rangle$ denotes dot product or cosine similarity. 
This formulation enables retrieval beyond exact term overlap, capturing paraphrases and deeper semantic relations. 
However, the query and document representations are tied to the embedding space of a specific model, making adaptation and transfer across different retrievers more challenging, leading to the model assumption.

\subsection{Pseudo-Relevance Feedback and Generative Relevance Feedback}
Pseudo-relevance feedback expands the initial query $q$ by leveraging the top-$k$ documents retrieved in the first stage. 
Let $\mathcal{D}_q^{(k)} = \{ d_1, d_2, \dots, d_k \}$ denotes the feedback set obtained from the initial retrieval.
Classical PRF methods estimate a relevance model $p(t \mid q)$ over terms $t \in \mathcal{V}$ using statistics from $\mathcal{D}_q^{(k)}$. 
A common formulation, as in RM3, interpolates the original query term distribution with the feedback model~\cite{abdul2004umass}:
\begin{equation}
    p(t \mid q’) = (1-\alpha) \cdot p(t \mid q) + \alpha 
    \cdot \sum_{d \in \mathcal{D}_q^{(k)}} p(t \mid d) \cdot p(d \mid q),
\end{equation}
where $q’$ is the reformulated query, $\alpha \in [0,1]$ controls the interpolation weight, and $p(t \mid \mathcal{D}_q^{(k)})$ is estimated from the feedback documents. 
This reformulation is then used for subsequent retrieval under the sparse retrieval framework.

In dense retrieval settings, PRF is performed directly in the embedding space.
VPRF refines the query representation $\textbf{q}$ by aggregating feedback document vectors~\cite{li2023pseudo,li2025query}:
\begin{equation}
    \mathbf{q}’ = \alpha \cdot \mathbf{q} + \beta \cdot \sum_{i=1}^k \mathbf{d}_i,
\end{equation}
where $\mathbf{d}_i = \text{Enc}(d_i), \, d_i \in \mathcal{D}_q^{(k)}$ is the embedding of the feedback document, while $\alpha, \beta$ control the contribution of the original query and feedback documents. 

On the other hand, large language models (LLMs) have recently been employed for query rewriting in natural language. 
Given the initial query, GRF uses an LLM parameterized by $\theta$ to construct expansions like pseudo-document, pseudo-answer, etc.~\cite{gao2023precise,shen2023large}: $d' \sim \text{LLM}_\theta(\mathcal{I}, \, q)$, where $\mathcal{I}$ denotes the instruction. 
It articulates the information needs in a more detailed form. The final reformulated query is obtained by concatenation:
\begin{equation}
    q' = [q; d'_1; d'_2; \dots],
\end{equation}
where $[\cdot ; \cdot ; \cdot]$ denotes text concatenation, and $d'_i$ represents different sample result. The reformulated query is then used for retrieval with either sparse or dense methods.

\section{Methodology}
\subsection{Generalized Pseudo-Relevance Feedback}
Building on both PRF and GRF, we propose \textit{Generalized Pseudo-Relevance Feedback} (GPRF), which integrates retrieval evidence with generative rewriting.
A comparison of GPRF against PRF, VPRF, and GRF is shown in Figure~\ref{fig:frame}.
Specifically, given the inital query $q$ and its top-$k$ retrieved documents $\mathcal{D}_q^{(k)}$, an LLM directly generates the rewritten query:
\begin{equation}
    q' \sim \text{LLM}_\theta(\mathcal{I}, \, q, \,   \mathcal{D}_q^{(k)}).
\end{equation}

\begin{figure*}[t]
    \centering
    \includegraphics[width=0.95\textwidth]{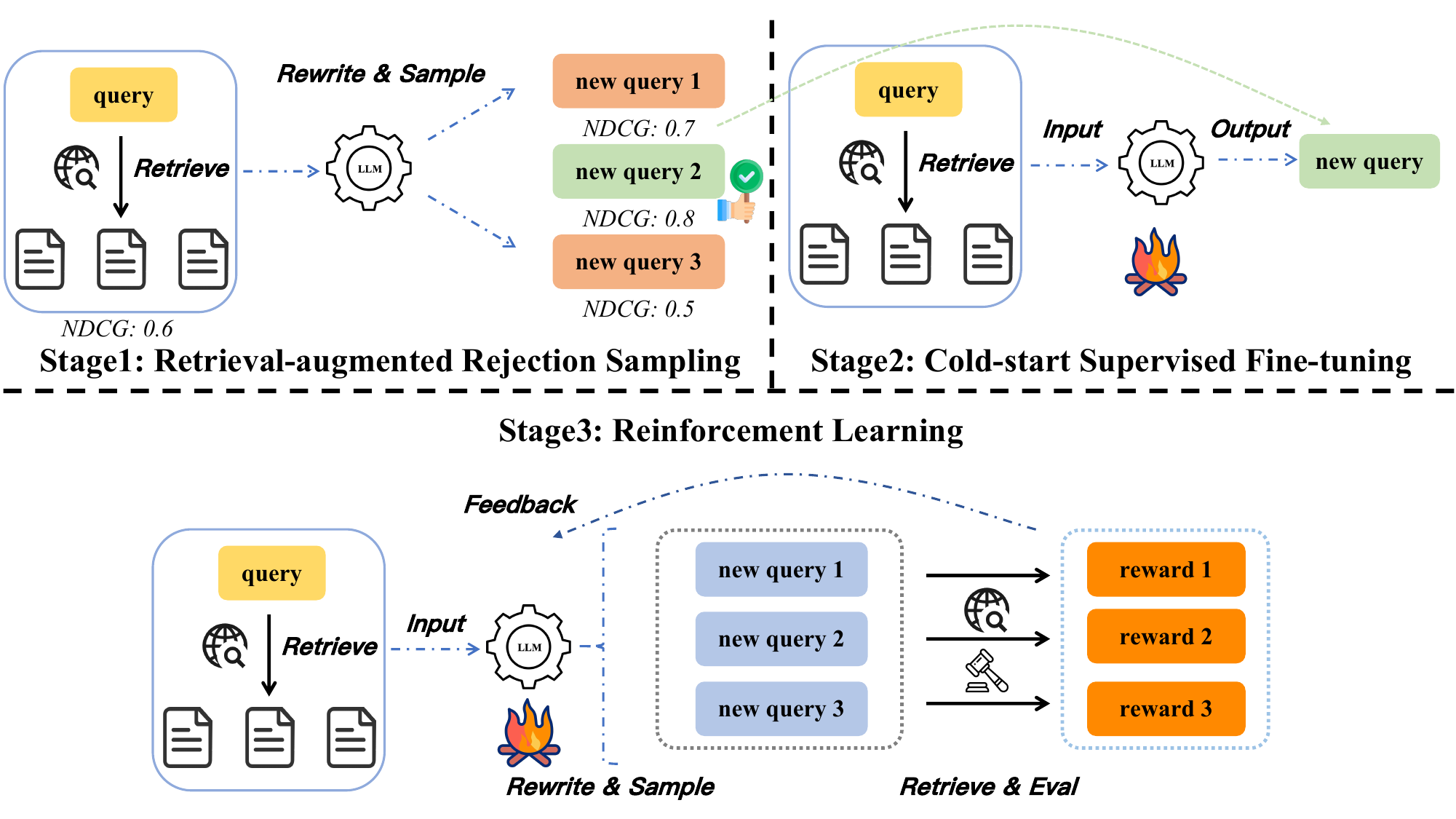}
    %\vspace{-4mm}
    \caption{Overview of the Utility-oriented Training Pipeline. The high-utility reformulations obtained via rejection sampling in Stage 1 are directly utilized as training labels for supervised fine-tuning (SFT) in Stage 2. In Stage 3, we directly use the performance on downstream retrieval tasks as the reward signal of reinforcement learning (RL).}
    \label{fig:method}
\end{figure*}

To enhance robustness and diversity, multiple reformulations can be sampled to capture different possible user intents.
These diverse reformulations can be conveniently incorporated into downstream retrieval, such as appending them directly to the original query (i.e., concatenation in Figure~\ref{fig:frame}), encoding them into embeddings and then aggregating them (just like PRF and VPRF), or directly conducting retrieval on these samples and post-processing the retrieval results.
This design combines the semantic grounding of PRF with the expressive generative capacity of LLMs, providing a model-agnostic mechanism that bridges sparse and dense retrieval.
On the other hand, GPRF can also be seamlessly combined with methods such as few-shot learning and Chain-of-Thought (CoT), which we leave for future work.

Nevertheless, this retrieval-augmented generation-based query rewriting paradigm is not without challenges.
Although grounding in feedback documents reduces hallucinations compared to GRF, their effectiveness is limited because the generative model remains sensitive to noisy or off-topic feedback documents, which may mislead reformulations and degrade retrieval performance~\cite{wang2024astute,tu2025robust}. 
These challenges motivate the development of a dedicated training method to control generation quality better and alleviate the influence of noisy feedback.

\subsection{Utility-oriented Training Pipeline}
\subsubsection{Overview} \hfill \break
As discussed above, a key challenge in generative query reformulation lies in the vulnerability of LLMs to noisy feedback documents: irrelevant or misleading evidence can easily distort the rewriting process, leading to suboptimal or even harmful expansions. 
To address this issue, we design a \textit{utility-oriented training pipeline} that explicitly incorporates ultimate retrieval performance into the model training process. 
By optimizing query rewriting not only for fluency or faithfulness, but also for retrieval effectiveness, the pipeline strengthens the model’s robustness against noisy inputs and enhances its ability to produce reliable, utility-driven reformulations aligned with the downstream retrieval task.

As shown in Figure~\ref{fig:method}, our pipeline consists of three stages. 
First, we perform sampling-based evaluation to identify the rewritten queries that maximize retrieval utility. 
Second, the best-performing samples are used to construct high-quality supervision signals for fine-tuning. 
Finally, reinforcement learning (RL) with direct utility-based rewards further aligns the model toward the ultimate goal of query rewriting. 
Together, these stages form an iterative framework that grounds query reformulation in retrieval performance while improving both accuracy and resilience.

\subsubsection{Retrieval-augmented Rejection Sampling} \hfill \label{subsec:rej_sample} \break 
The first stage of our pipeline is retrieval-augmented rejection sampling, which aims to filter out low-utility query reformulations and retain only those that improve retrieval effectiveness.
Concretely, given an initial query $q$ and its top-$k$ feedback documents $\mathcal{D}_q^{(k)}$, the rewriting model ($\text{LLM}_\theta$) generates a set of candidate reformulations:
\begin{equation}
\{ q'_1, q'_2, \dots, q'_{M} \} \sim \text{LLM}_{\theta}(\mathcal{I},  \, q, \, \mathcal{D}_q^{(k)}), 
\end{equation}
where $M$ denotes the number of sampled rewrites (e.g., $M=10$) and $\mathcal{I}$ is the instruction (detailed in Table~\ref{tab:prompt}).
Each reformulated query $q'_j$ is then submitted back to the retrieval system, producing a ranked list of documents $\mathcal{D}_{q'_j}^{(k)}$. To evaluate its effectiveness, we measure a utility function $U(\cdot)$, defined as the improvement in retrieval quality (for example, NDCG@10~\cite{jarvelin2002cumulated}) compared to the original query $q$:
\begin{equation}
    U(q'_j) = \text{NDCG@10}(\mathcal{D}_{q'_j}^{(k)}) - \text{NDCG@10}(\mathcal{D}_{q}^{(k)}).
\end{equation}
Finally, we select the reformulation with the highest utility score as the accepted rewrite:
\begin{equation}
    q^{\ast} = \arg \max_{q'_j} U(q'_j).
\end{equation}

\begin{table}[t]
\centering
\caption{The prompt template for the query rewriting module of GPRF. For the various tasks and datasets presented in this paper, we employ a uniform prompt.}
\begin{minipage}{0.95\linewidth}
\begin{tcolorbox}[colback=gray!5, colframe=black!40,listing only, listing options={basicstyle=\ttfamily\small,breaklines}]
Please rewrite the user's query based on several relevant passages (which may contain noise or errors). The rewritten query should preserve the original meaning while incorporating as much information as possible, so that search engines can more effectively retrieve relevant passages.

Relevant Passages:

Passage 1: \{passage 1\}

Passage 2: \{passage 2\}

......

User Query: \{question\}

Rewritten Query:
\end{tcolorbox}
\end{minipage}
\label{tab:prompt}
\end{table}

This procedure ensures that only rewrites yielding the greatest retrieval improvement are retained, while others are discarded.
In this way, retrieval-augmented rejection sampling provides high-quality pseudo-supervision signals for the subsequent supervised fine-tuning stage, effectively grounding query rewriting in retrieval performance and mitigating the influence of noisy feedback.

\subsubsection{Cold-start Supervised Fine-tuning (SFT)} \hfill \break
After obtaining high-utility reformulations from retrieval-augmented rejection sampling, we use them as pseudo-supervision signals to initialize the rewriting model. 
This stage provides the model with explicit guidance on how to generate reformulations that improve retrieval performance.

Formally, let $D_{\text{SFT}} = \{(x, y)\}$ denotes the SFT dataset, where input $x = (\mathcal{I}; q; \mathcal{D}_{q}^{(k)})$ can be regarded as a combination of the instruction $\mathcal{I}$, the initial query $q$, and the original feedback $\mathcal{D}_{q}^{(k)}$. The output $y = q^{\ast}$ is the selected reformulation with the highest utility for query $q$. 
It minimizes the negative log-likelihood (NLL) of generating the target reformulation:
\begin{equation}
\mathcal{L}_{\text{SFT}}(\theta) = -\sum_{i=1}^{|y|} \log p_\theta(y_i \mid x, \, y_{<i}).
\end{equation}
This training step encourages the model to imitate utility-driven reformulations, thereby reducing its tendency to be misled by noisy feedback. 
By grounding the model in supervised signals, SFT establishes a strong initialization that enhances both stability and convergence in the subsequent RL stage.

\subsubsection{Reinforcement Learning (RL)} \hfill \break
While SFT provides the model with high-utility reformulation examples, it cannot fully address the variability of real retrieval scenarios, where feedback may be noisy and ambiguous.
To further align the model with retrieval-oriented objectives, we adopt reinforcement learning (RL) with the Generalized Reweighted Policy Optimization (GRPO)~\cite{guo2025deepseek,shao2024deepseekmath} algorithm.
It samples a group of outputs $G = \{y_1, \dots, y_{|G|}\}$ for each input $x$ during training, and each $y_i$ corresponds to a reward $r_i$.
To jointly account for retrieval performance at top ranks and the overall recall, we adopt a multi-view reward function defined as:
\begin{equation}
    r_i = \text{NDCG@10}(\mathcal{D}_{y_i}^{(k)}) + \lambda \cdot \text{Recall@100}(\mathcal{D}_{y_i}^{(k)}),
\end{equation}
where $\lambda$ is a trade-off hyper-parameter. The rewards are then normalized within the group to produce the advantage function:
\begin{equation}
\hat{A}_{i} = \frac{r_i - \text{mean}\left(\left\{r_1,\dots,r_{|G|}\right\}\right)}{\text{std}\left(\left\{r_1,\dots,r_{|G|}\right\}\right)}
\end{equation}

Thus, the overall loss function is formulated as:
\begin{equation}\begin{aligned}
\mathcal{L}_{\text{GRPO}}(\theta) = &
-\frac{1}{|G|}\sum_{i=1}^{|G|}\frac{1}{|y_i|}\sum_{t=1}^{|y_i|} \min 
\bigg(r_{i,t}(\theta)\hat{A}_{i}, \\
& \text{clip}\left(r_{i,t}(\theta), 1- \epsilon, 1 + \epsilon \right) \hat{A}_{i} \bigg) 
- \beta D_{\text{KL}} \left[\pi_\theta || \pi_{\text{ref}}\right],
\end{aligned}\end{equation}
where $r_{i,t}(\theta) = \frac{\pi_\theta(y_{i, t} \mid x, y_{i, <t})}{\pi_{\text{old}}(y_{i, t} \mid x, y_{i, <t})}$ is the importance ratio, and $\epsilon$ as well as $\beta$ are hyper-parameters.
Through this RL stage, the model is directly optimized for retrieval effectiveness rather than imitation alone. 
Combined with rejection sampling and SFT, GRPO equips the rewriting model with greater robustness to noisy feedback and stronger utility-driven reformulation capabilities. 
\section{Experimental Setup}
\subsection{Datasets \& Evaluation Metrics}
To train the GPRF model, we use the MS-MARCO Passage Retrieval dataset~\cite{bajaj2016ms}, which provides large-scale query–document pairs for supervised retrieval.
In the cold-start SFT stage, we sample 200k instances from the dataset and apply rejection sampling (following \S\ref{subsec:rej_sample}) based on a commonly used sparse retriever, BM25~\cite{robertson2009probabilistic}, and a dense retriever, e5-base-v2~\cite{wang2022text}, selecting the top 30k instances with the greatest improvement from both retrievers to construct the training set $D_{\text{SFT}}$. 
In the RL stage, we similarly sample 200k instances directly from the dataset, while randomly assigning BM25 or e5-base-v2 as the retriever to construct the reward function.

We evaluate the performance of the retrieval model before and after query rewriting on both in-domain and out-of-domain retrieval benchmarks to assess the effectiveness and generalizability comprehensively.
For in-domain evaluation, we report results on the MS-MARCO Passage Retrieval dev set (MS dev)~\cite{bajaj2016ms} as well as the TREC Deep Learning track 2019 (DL19) and 2020 (DL20)~\cite{craswell2025overview}.
To evaluate out-of-domain robustness, following \citet{gao2023precise}, we test on six publicly available low-resource datasets from the BEIR benchmark~\cite{thakur2021beir}, namely ArguAna, DBPedia, FiQA-2018, SCIDOCS, SciFact, and TREC-COVID.
Following previous works~\cite{gao2023precise,li2025llm}, we adopt two standard retrieval metrics: NDCG@10 and Recall@100 (R@100). 
NDCG@10 emphasizes effectiveness for highly relevant documents at top ranks, while R@100 reflects the system’s ability to cover a broader set of relevant results.

\subsection{Baselines}
%We compare GPRF against a diverse set of baselines, covering both classical sparse retrieval and neural dense retrieval models.
%On top of these retrievers, 
We mainly compare GPRF with three categories of methods.
The first is the direct retrieval baseline, where no rewriting is applied and the system relies solely on the original query. 
The second category includes PRF–based approaches. 
For sparse retrieval, we use RM3~\cite{abdul2004umass}, a classical lexical feedback method that expands the query distribution with terms from top-ranked documents. 
For dense retrieval, we adopt VPRF~\cite{li2023pseudo}, which refines the query embedding by aggregating representations of feedback documents. 
The third category consists of zero-shot GRF methods that employ LLMs for query rewriting. 
Specifically, we consider three methods: 
HyDE~\cite{gao2023precise}, which generates hypothetical answer passages as pseudo-documents to enrich queries;
CoT~\cite{jagerman2023query}, which leverages the Chain-of-Thought to provide a rationale for the pseudo-answer;
and LameR~\cite{shen2023large}, which follows a retrieve–answer–retrieve pipeline where pseudo-answers are generated to improve retrieval performance.
These approaches provide a comprehensive comparison, allowing us to evaluate GPRF not only against traditional lexical and dense feedback methods but also against recent LLM-based generative rewriting approaches under both in-domain and out-of-domain retrieval settings.

\subsection{Implementation Details}
\subsubsection{Model Selection}
We experiment with various retrieval models as well as backbone LLMs.
For retrievers, we consider BM25~\cite{robertson2009probabilistic} as the classical sparse approach, along with two dense retrievers: e5-base-v2 (E5)~\cite{wang2022text}, which serves as our primary in-domain dense
retriever, and bge-base-en-v1.5 (BGE)~\cite{xiao2024c}, which is not employed during the training stage and therefore functions as an out-of-domain model to test generalizability.
For the query rewriting model, to balance the performance and efficiency, we select two LLMs of moderate size: Llama-3.2-3B-Instruct~\cite{dubey2024llama} (Llama) and Qwen2.5-3B-Instruct~\cite{bai2023qwen} (Qwen).

\subsubsection{Training Settings}
We use four \textit{NVIDIA A100-SXM4-40GB} GPUs for training GPRF models. 
In the SFT stage, the model is trained for 2 epochs with a learning rate of 1e-6. 
We set both the \textit{per-device training batch size} and the \textit{gradient accumulation steps} to 8. 
In the RL stage, we train the model for 1 epoch with the same learning rate.
Here, we increase the \textit{per-device batch size} and \textit{gradient accumulation steps} to 16, set the group size $|G|=8$, use a sampling temperature of 1.0, and apply the KL-divergence regularization term $\beta$ with 1e-3.

\subsubsection{Evaluation Settings}
For evaluation, we set the temperature to 0 to ensure deterministic decoding, feed $k=10$ retrieved documents into the LLMs, and sample $M=10$ reformulated queries for each input. 
To combine these reformulations with retrieval, we follow different strategies for sparse and dense retrievers. 
For BM25, we concatenate all reformulated queries with the original query to form the final input. 
For dense retrievers, we apply the VPRF strategy instead, aggregating the embeddings of the reformulated queries to construct the refined query representation.

\section{Results and Analysis}
\begin{table}[t]
\renewcommand{\arraystretch}{1.3}
\centering
\caption{Evaluation results of different rewriting methods on in-domain datasets. The best and second-best methods of each retriever are marked in bold and underlined, respectively. "L" and "Q" denote using Llama-3-3.2B-Instruct and Qwen2.5-3B-Instruct as the backbone model, while "$\dagger$" and "$\ddagger$" indicate significantly worse than the best and second-best method at the $p < 0.05$ level using the two-tailed pairwise t-test, respectively.}
\label{tab:ind}
\resizebox{0.99\columnwidth}{!}{
\begin{tabular}{c||cc|cc|cc}
\toprule
\multirow{2}{*}{Method} & \multicolumn{2}{c|}{MS dev} & \multicolumn{2}{c|}{DL 19} & \multicolumn{2}{c}{DL 20} \\
\cline{2-7} 
 & NDCG@10 & R@100 & NDCG@10 & R@100 & NDCG@10 & R@100 \\ 
\midrule
BM25 & 0.2284$^{\dagger\ddagger}$ & 0.6578$^{\dagger\ddagger}$ & 0.5058$^{\dagger\ddagger}$ & 0.4531$^{\dagger\ddagger}$ & 0.4796$^{\dagger\ddagger}$ & 0.4834$^{\dagger\ddagger}$ \\
+RM3 & 0.2023$^{\dagger\ddagger}$ & 0.6538$^{\dagger\ddagger}$ & 0.5216$^{\dagger\ddagger}$ & 0.4821$^{\dagger}$ & 0.4896$^{\dagger\ddagger}$ & 0.5316$^{\dagger}$ \\
+HyDE$_L$ & 0.2023$^{\dagger\ddagger}$ & 0.6425$^{\dagger\ddagger}$ & 0.6001$^{\dagger}$ & 0.4795$^{\dagger}$ & 0.5733$^{\dagger\ddagger}$ & 0.5542 \\
+HyDE$_Q$ & 0.2224$^{\dagger\ddagger}$ & 0.6829$^{\dagger\ddagger}$ & 0.6030$^{\dagger}$ & 0.4890$^{\dagger}$ & 0.5845$^{\dagger\ddagger}$ & 0.5539 \\
+CoT$_L$ & 0.2233$^{\dagger\ddagger}$ & 0.6786$^{\dagger\ddagger}$ & 0.6215$^{\dagger}$ & 0.4923 & 0.5973$^{\dagger\ddagger}$ & \underline{0.5738} \\
+CoT$_Q$ & 0.2339$^{\dagger\ddagger}$ & 0.6914$^{\dagger\ddagger}$ & 0.5480$^{\dagger\ddagger}$ & 0.4625$^{\dagger}$ & 0.5468$^{\dagger\ddagger}$ & 0.5535 \\
+Lamer$_L$ & 0.2367$^{\dagger\ddagger}$ & 0.6773$^{\dagger\ddagger}$ & 0.6361$^{\dagger}$ & 0.4849$^{\dagger}$ & 0.5975$^{\dagger}$ & 0.5718 \\
+Lamer$_Q$ & 0.2593$^{\dagger\ddagger}$ & 0.6830$^{\dagger\ddagger}$ & \underline{0.6589} & \underline{0.5091} & 0.6219 & 0.5594 \\
+GPRF$_L$ & \textbf{0.3208} & \textbf{0.7486} & \textbf{0.6917} & \textbf{0.5401} & \textbf{0.6707} & \textbf{0.5849} \\
+GPRF$_Q$ & \underline{0.3016}$^{\dagger}$ & \underline{0.7179}$^{\dagger}$ & 0.6461$^{\dagger}$ & 0.4952 & \underline{0.6332} & 0.5343 \\

\midrule

E5 & 0.4179$^{\dagger\ddagger}$ & \underline{0.8878}$^{\dagger}$ & 0.7048 & 0.5375 & 0.7039$^{\dagger\ddagger}$ & 0.6019$^{\dagger\ddagger}$  \\
+VPRF & 0.3262$^{\dagger\ddagger}$ & 0.8555$^{\dagger\ddagger}$ & 0.6765$^{\dagger}$ & \textbf{0.5671} & 0.7027$^{\dagger\ddagger}$ & 0.5943$^{\dagger\ddagger}$  \\
+HyDE$_L$ & 0.3291$^{\dagger\ddagger}$ & 0.8124$^{\dagger\ddagger}$ & 0.7096 & 0.5273$^{\dagger}$ & 0.6895$^{\dagger\ddagger}$ & 0.5871$^{\dagger\ddagger}$  \\
+HyDE$_Q$ & 0.3579$^{\dagger\ddagger}$ & 0.8467$^{\dagger\ddagger}$ & 0.6781 & 0.5344 & 0.7006$^{\dagger\ddagger}$ & 0.6005$^{\dagger\ddagger}$  \\
+CoT$_L$ & 0.3036$^{\dagger\ddagger}$ & 0.7755$^{\dagger\ddagger}$ & 0.5992$^{\dagger\ddagger}$ & 0.4532$^{\dagger\ddagger}$ & 0.6001$^{\dagger\ddagger}$ & 0.5218$^{\dagger\ddagger}$  \\
+CoT$_Q$ & 0.2983$^{\dagger\ddagger}$ & 0.7767$^{\dagger\ddagger}$ & 0.5941$^{\dagger\ddagger}$ & 0.4507$^{\dagger\ddagger}$ & 0.5988$^{\dagger\ddagger}$ & 0.5088$^{\dagger\ddagger}$  \\
+Lamer$_L$ & 0.3459$^{\dagger\ddagger}$ & 0.8046$^{\dagger\ddagger}$ & 0.6723$^{\dagger}$ & 0.4968$^{\dagger\ddagger}$ & 0.7096$^{\dagger\ddagger}$ & 0.5955$^{\dagger\ddagger}$  \\
+Lamer$_Q$ & 0.3594$^{\dagger\ddagger}$ & 0.8011$^{\dagger\ddagger}$ & 0.6873$^{\dagger}$ & 0.4881$^{\dagger\ddagger}$ & 0.7297$^{\dagger}$ & 0.6094$^{\dagger}$  \\
+GPRF$_L$ & \textbf{0.4283} & 0.8852$^{\dagger}$ & \underline{0.7228} & 0.5405 & \textbf{0.7585} & \underline{0.6205} \\
+GPRF$_Q$ & \underline{0.4231}$^{\dagger}$ & \textbf{0.8904} & \textbf{0.7382} & \underline{0.5541} & \underline{0.7524} & \textbf{0.6257} \\

\midrule

BGE & 0.4134$^{\dagger\ddagger}$ & \underline{0.8856}$^{\dagger}$ & 0.7245 & 0.5174$^{\dagger}$ & 0.7052$^{\dagger\ddagger}$ & 0.5797 \\
+VPRF & 0.3200$^{\dagger\ddagger}$ & 0.8500$^{\dagger\ddagger}$ & 0.7096$^{\dagger}$ & 0.5436 & 0.6921$^{\dagger\ddagger}$ & 0.5765$^{\dagger}$ \\
+HyDE$_L$ & 0.3348$^{\dagger\ddagger}$ & 0.8197$^{\dagger\ddagger}$ & 0.7263 & 0.5390$^{\dagger}$ & 0.7231 & 0.5769 \\
+HyDE$_Q$ & 0.3639$^{\dagger\ddagger}$ & 0.8527$^{\dagger\ddagger}$ & 0.7002$^{\dagger}$ & 0.5388 & 0.7197$^{\dagger}$ & 0.5720$^{\dagger}$ \\
+CoT$_L$ & 0.3531$^{\dagger\ddagger}$ & 0.8288$^{\dagger\ddagger}$ & 0.6898$^{\dagger\ddagger}$ & 0.4935$^{\dagger\ddagger}$ & 0.6822$^{\dagger\ddagger}$ & 0.5572$^{\dagger}$ \\
+CoT$_Q$ & 0.3407$^{\dagger\ddagger}$ & 0.8224$^{\dagger\ddagger}$ & 0.6738$^{\dagger\ddagger}$ & 0.5036$^{\dagger\ddagger}$ & 0.6363$^{\dagger\ddagger}$ & 0.5101$^{\dagger\ddagger}$ \\
+Lamer$_L$ & 0.3676$^{\dagger\ddagger}$ & 0.8368$^{\dagger\ddagger}$ & 0.7495 & \underline{0.5587} & 0.7210$^{\dagger}$ & \underline{0.5929} \\
+Lamer$_Q$ & 0.3754$^{\dagger\ddagger}$ & 0.8407$^{\dagger\ddagger}$ & \underline{0.7581} & 0.5341$^{\dagger}$ & 0.7263 & 0.5441$^{\dagger\ddagger}$ \\
+GPRF$_L$ & \textbf{0.4262} & 0.8846$^{\dagger}$ & 0.7555 & 0.5560 & \underline{0.7384} & 0.5778 \\
+GPRF$_Q$ & \underline{0.4234} & \textbf{0.8897} & \textbf{0.7612} & \textbf{0.5711} & \textbf{0.7613} & \textbf{0.6025} \\
\bottomrule
\end{tabular}
}
%\vspace{-4mm}
\end{table}

In this section, we mainly aim to explore the following three research questions thoroughly:
\begin{itemize}[leftmargin=*]
\item \textbf{RQ1:} Can GPRF perform effectively on in-domain data and generalize to out-of-domain data at the same time?
\item \textbf{RQ2:} Can GPRF relax the relevance assumption and tolerate noisy data in feedback documents?
\item \textbf{RQ3:} Can GPRF mitigate the model assumption and perform effectively for retrievers not presented in the training process, with top documents retrieved by or not by themselves? 
\end{itemize}

\subsection{Main Results (RQ1)}
\begin{table*}[htbp]
\centering
\caption{Evaluation results of different rewriting methods on out-of-domain datasets with Llama as the backbone model. The best and second-best methods of each retriever are marked in bold and underlined, respectively. "$\dagger$" and "$\ddagger$" indicate significantly worse than the best and second-best method at the $p < 0.05$ level using the two-tailed pairwise t-test, respectively.}
\label{tab:ood}
\renewcommand{\arraystretch}{1.3}
\resizebox{0.99\textwidth}{!}{
\begin{tabular}{c||cc|cc|cc|cc|cc|cc||cc}
\toprule
 & \multicolumn{2}{c|}{ArguAna} & \multicolumn{2}{c|}{DBPedia} & \multicolumn{2}{c|}{FiQA-2018} & \multicolumn{2}{c|}{SCIDOCS} & \multicolumn{2}{c|}{SciFact} & \multicolumn{2}{c||}{TREC-COVID} & \multicolumn{2}{c}{Avg.} \\
\multirow{-2}{*}{Method} & NDCG@10 & R@100 & NDCG@10 & R@100 & NDCG@10 & R@100 & NDCG@10 & R100 & NDCG@10 & R@100 & NDCG@10 & R@100 & NDCG@10 & R@100 \\ 
\midrule

BM25 & 0.2999$^{\dagger}$ & 0.9324$^{\dagger\ddagger}$ & 0.3180$^{\dagger\ddagger}$ & 0.4682$^{\dagger\ddagger}$ & \underline{0.2361}$^{\dagger}$ & \underline{0.5395}$^{\dagger}$ & 0.1490$^{\dagger}$ & 0.3477$^{\dagger\ddagger}$ & 0.6789$^{\dagger}$ & 0.9253 & 0.5947$^{\dagger\ddagger}$ & 0.1091$^{\dagger\ddagger}$ & 0.3794 & 0.5537 \\
+RM3 & 0.2865$^{\dagger\ddagger}$ & \textbf{0.9552} & 0.3080$^{\dagger\ddagger}$ & 0.4594$^{\dagger\ddagger}$ & 0.1916$^{\dagger\ddagger}$ & 0.4967$^{\dagger\ddagger}$ & \underline{0.1491}$^{\dagger}$ & \underline{0.3621} & 0.6457$^{\dagger\ddagger}$ & 0.9147$^{\dagger}$ & 0.5927$^{\dagger\ddagger}$ & 0.1168$^{\dagger}$ & 0.3623 & 0.5508 \\
+HyDE & 0.2794$^{\dagger\ddagger}$ & 0.9260$^{\dagger\ddagger}$ & 0.3303$^{\dagger\ddagger}$ & 0.4719$^{\dagger\ddagger}$ & 0.1803$^{\dagger\ddagger}$ & 0.4931$^{\dagger\ddagger}$ & 0.1140$^{\dagger\ddagger}$ & 0.3283$^{\dagger\ddagger}$ & 0.6557$^{\dagger\ddagger}$ & 0.9377 & 0.6458$^{\dagger}$ & 0.1264$^{\dagger}$ & 0.3676 & 0.5472 \\
+CoT & \underline{0.3013}$^{\dagger}$ & 0.9403$^{\dagger}$ & 0.3510$^{\dagger\ddagger}$ & 0.5054 & 0.2081$^{\dagger\ddagger}$ & 0.5187$^{\dagger}$ & 0.1372$^{\dagger\ddagger}$ & 0.3532$^{\dagger}$ & 0.6971 & \textbf{0.9437} & 0.6815$^{\dagger}$ & 0.1264$^{\dagger}$ & \underline{0.3960} & \underline{0.5646} \\
+Lamer & 0.2547$^{\dagger\ddagger}$ & 0.9189$^{\dagger\ddagger}$ & \underline{0.3909} & \underline{0.5164} & 0.1970$^{\dagger\ddagger}$ & 0.4703$^{\dagger\ddagger}$ & 0.1267$^{\dagger\ddagger}$ & 0.3120$^{\dagger\ddagger}$ & \underline{0.7047} & \underline{0.9397} & \underline{0.6824}$^{\dagger}$ & \underline{0.1298}$^{\dagger}$ & 0.3927 & 0.5479 \\
+GPRF & \textbf{0.3139} & \underline{0.9452} & \textbf{0.4009} & \textbf{0.5200} & \textbf{0.2912} & \textbf{0.5935} & \textbf{0.1579} & \textbf{0.3665} & \textbf{0.7127} & 0.9367 & \textbf{0.7738} & \textbf{0.1441} & \textbf{0.4417} & \textbf{0.5843} \\

\midrule

E5 & 0.3258$^{\dagger\ddagger}$ & 0.9467$^{\dagger\ddagger}$ & \underline{0.4226}$^{\dagger}$ & \textbf{0.5420} & \underline{0.3991}$^{\dagger}$ & 0.7324$^{\dagger}$ & \underline{0.1862}$^{\dagger}$ & \underline{0.4211} & 0.7200$^{\dagger\ddagger}$ & \textbf{0.9627} & 0.6961$^{\dagger\ddagger}$  & 0.1287$^{\dagger\ddagger}$  & 0.4583 & 0.6223 \\
+VPRF & \textbf{0.3519} & \textbf{0.9630} & 0.3866$^{\dagger\ddagger}$ & 0.5060$^{\dagger\ddagger}$ & 0.2974$^{\dagger\ddagger}$ & 0.6839$^{\dagger\ddagger}$ & 0.1749$^{\dagger\ddagger}$ & 0.4181 & 0.5798$^{\dagger\ddagger}$ & 0.9467 & 0.6923$^{\dagger\ddagger}$  & 0.1203$^{\dagger\ddagger}$  & 0.4138 & 0.6063 \\
+HyDE & 0.3251$^{\dagger}$ & \underline{0.9573}$^{\dagger}$ & 0.4020$^{\dagger\ddagger}$ & 0.5259 & 0.3818$^{\dagger}$ & \underline{0.7336} & 0.1833$^{\dagger}$ & 0.4181 & \textbf{0.7465} & \textbf{0.9627} & \underline{0.7614} & \underline{0.1392} & \underline{0.4667} & \underline{0.6228} \\
+CoT & 0.3236$^{\dagger\ddagger}$ & 0.9467$^{\dagger\ddagger}$ & 0.3785$^{\dagger\ddagger}$ & 0.4755$^{\dagger\ddagger}$ & 0.3728$^{\dagger\ddagger}$ & 0.7081$^{\dagger\ddagger}$ & 0.1675$^{\dagger\ddagger}$ & 0.3927$^{\dagger\ddagger}$ & 0.7012$^{\dagger\ddagger}$ & \underline{0.9593} & 0.5710$^{\dagger\ddagger}$  & 0.0982$^{\dagger\ddagger}$  & 0.4191 & 0.5968 \\
+Lamer & 0.3086$^{\dagger\ddagger}$ & 0.9410$^{\dagger\ddagger}$ & 0.4032$^{\dagger\ddagger}$ & 0.4682$^{\dagger\ddagger}$ & 0.3868$^{\dagger}$ & 0.7230$^{\dagger}$ & 0.1739$^{\dagger\ddagger}$ & 0.3987$^{\dagger\ddagger}$ & 0.7046$^{\dagger\ddagger}$ & 0.9523 & 0.6708$^{\dagger\ddagger}$  & 0.1117$^{\dagger\ddagger}$  & 0.4413 & 0.5992 \\
+GPRF & \underline{0.3285}$^{\dagger}$ & 0.9481$^{\dagger}$ & \textbf{0.4442} & \underline{0.5355} & \textbf{0.4323} & \textbf{0.7469} & \textbf{0.1893} & \textbf{0.4239} & \underline{0.7404} & \underline{0.9593} & \textbf{0.7642} & \textbf{0.1396} & \textbf{0.4832} & \textbf{0.6256} \\

\midrule

BGE & \underline{0.4534} & \underline{0.9915} & 0.4078$^{\dagger}$ & \underline{0.5301} & \underline{0.4064} & \textbf{0.7415} & \underline{0.2168} & 0.4957 & 0.7394 & 0.9633$^{\dagger}$ & 0.7802$^{\dagger}$ & 0.1407 & \underline{0.5007} & 0.6438 \\
+VPRF & 0.4383$^{\dagger\ddagger}$ & 0.9908 & 0.3805$^{\dagger\ddagger}$ & 0.5009$^{\dagger\ddagger}$ & 0.2898$^{\dagger\ddagger}$ & 0.6601$^{\dagger\ddagger}$ & 0.2037$^{\dagger\ddagger}$ & 0.4905$^{\dagger}$ & 0.6080$^{\dagger\ddagger}$ & 0.9433$^{\dagger\ddagger}$ & \textbf{0.8204} & \textbf{0.1489} & 0.4568 & 0.6224 \\
+HyDE & 0.4156$^{\dagger\ddagger}$ & 0.9872$^{\dagger}$ & 0.4052$^{\dagger\ddagger}$ & 0.5284 & 0.3919$^{\dagger}$ & 0.7338 & 0.2148 & \textbf{0.5003} & \textbf{0.7488} & \textbf{0.9767} & \underline{0.8056} & \underline{0.1481} & 0.4970 & \underline{0.6457} \\
+CoT & 0.4520 & 0.9900 & 0.3953$^{\dagger\ddagger}$ & 0.5065$^{\dagger\ddagger}$ & 0.3841$^{\dagger\ddagger}$ & 0.7123$^{\dagger\ddagger}$ & 0.2001$^{\dagger\ddagger}$ & 0.4656$^{\dagger\ddagger}$ & 0.7395 & 0.9600$^{\dagger}$ & 0.7530$^{\dagger\ddagger}$ & 0.1347$^{\dagger\ddagger}$ & 0.4873 & 0.6282 \\
+Lamer & 0.4426$^{\dagger\ddagger}$ & 0.9900 & \underline{0.4101}$^{\dagger}$ & 0.4922$^{\dagger\ddagger}$ & 0.3751$^{\dagger\ddagger}$ & 0.7221$^{\dagger}$ & 0.2158 & 0.4778$^{\dagger\ddagger}$ & 0.7428 & 0.9617 & 0.7856 & 0.1394$^{\dagger\ddagger}$ & 0.4953 & 0.6305 \\
+GPRF & \textbf{0.4542} & \textbf{0.9922} & \textbf{0.4285} & \textbf{0.5393} & \textbf{0.4119} & \underline{0.7395} & \textbf{0.2198} & \underline{0.4960} & \underline{0.7482} & \underline{0.9700} & 0.7909 & 0.1430 & \textbf{0.5089} & \textbf{0.6467} \\
\bottomrule
\end{tabular}
}
%\vspace{-2mm}
\end{table*}
Table~\ref{tab:ind} presents the evaluation results on three in-domain datasets. 
We observe that our GPRF consistently outperforms all baselines across different retrievers and evaluation metrics.
For the sparse retriever BM25, both RM3 and recent GRF methods (i.e., HyDE, CoT, LameR) bring moderate gains, but their improvements are unstable and often limited by the relevance assumption. 
In contrast, GPRF achieves substantial improvements, with up to 40.5\% NDCG@10 improvement on MS dev and 39.8\% on DL 20 compared to the vanilla BM25 retriever, demonstrating its strong ability to leverage retrieval evidence and generate effective reformulations.
For dense retrievers, E5 and BGE, similar trends can be observed. 
While VPRF provides some benefits by aggregating document embeddings, its performance lags behind GRF methods that operate at the natural language level and break the constraints of model assumptions. 
Among GRF baselines, LameR and HyDE are competitive, but GPRF consistently delivers the best or second-best results across nearly all settings. 
Notably, on DL20 with BGE, GPRF boosts NDCG@10 from 0.7052 to 0.7613, significantly surpassing the best performance of other baselines.
When comparing different backbone LLMs, both Llama and Qwen yield robust results, with Qwen generally better suited to BGE and Llama more aligned with BM25. 
For E5, Qwen demonstrates stronger recall performance, while Llama delivers slightly better precision-oriented gains (NDCG@10).
This suggests that GPRF is not only effective but also adaptable across different model backbones.
Overall, it demonstrates that GPRF is particularly advantageous in in-domain settings, consistently mitigating the weaknesses of both traditional PRF and GRF, delivering substantial and statistically significant improvements across varying sparse and dense retrievers.

\begin{table}[t]
\renewcommand{\arraystretch}{1.3}
\centering
\caption{Ablation study on the impact of different training stages with Llama. The best and second-best methods of each retriever are marked in bold and underlined, respectively. "Vanilla" denotes that the model is not trained.}
\label{tab:abl}
\resizebox{0.99\columnwidth}{!}{
\begin{tabular}{c|c||cc|cc|cc}
\toprule
\multirow{2}{*}{Retriever} & \multirow{2}{*}{Method} & \multicolumn{2}{c|}{MS dev} & \multicolumn{2}{c|}{DL 19} & \multicolumn{2}{c}{DL 20} \\
\cline{3-8} 
 & &  NDCG@10 & R@100 & NDCG@10 & R@100 & NDCG@10 & R@100 \\ 
 \midrule
\multirow{4}{*}{BM25} 
& Vanilla & 0.2360 & 0.6651 & 0.6182 & 0.4964 & 0.5751 & 0.5624\\
& SFT-only & 0.2511 & 0.6726 & 0.6280 & 0.4765 & 0.5890 & 0.5542 \\
& RL-only & \underline{0.3061} & \underline{0.7382} & \underline{0.6598} & \underline{0.5389} & \underline{0.6480} & \underline{0.5689} \\
& GPRF & \textbf{0.3208} & \textbf{0.7486} & \textbf{0.6917} & \textbf{0.5401} & \textbf{0.6707} & \textbf{0.5849} \\
\midrule
\multirow{4}{*}{E5} 
& Vanilla & 0.3361 & 0.7979 & 0.6631 & 0.4780 & 0.6384 & 0.5316 \\
& SFT-only & 0.3677 & 0.8530 & 0.7183 & \underline{0.5380} & 0.6978 & 0.5948 \\
& RL-only & \underline{0.4219} & \underline{0.8806} & \underline{0.7209} & 0.5373 & \underline{0.7432} & \underline{0.6145}   \\
& GPRF & \textbf{0.4283} & \textbf{0.8852} & \textbf{0.7228} & \textbf{0.5405} & \textbf{0.7585} & \textbf{0.6205} \\
\midrule
\multirow{4}{*}{BGE} 
& Vanilla & 0.3693 & 0.8496 & 0.7513 & 0.5488 & 0.7293 & 0.5719 \\
& SFT-only & 0.3842 & 0.8676 & \underline{0.7523} & \textbf{0.5619} & \underline{0.7375} & \textbf{0.5948} \\
& RL-only & \underline{0.4183} & \underline{0.8763} & 0.7418 & 0.5328 & 0.7222 & 0.5635 \\
& GPRF & \textbf{0.4262} & \textbf{0.8846} & \textbf{0.7555} & \underline{0.5560} & \textbf{0.7384} & \underline{0.5778} \\
\bottomrule
\end{tabular}
}
%\vspace{-4mm}
\end{table}

Table~\ref{tab:ood} further shows the results on six out-of-domain datasets when using Llama as the backbone rewriting model.
It can be observed that GPRF still achieves the best overall performance across all retrievers and datasets, consistently outperforming both classical PRF and recent GRF approaches.
For the sparse retriever BM25, traditional PRF (i.e., RM3) often fails to generalize and even underperforms the vanilla BM25 algorithm on several datasets, confirming its vulnerability to noisy feedback in distribution-shifted settings.
GRF methods such as HyDE, CoT, and LameR yield moderate improvements on specific datasets (e.g., CoT on SciFact, LameR on DBPedia), but their performance is highly inconsistent and fails to dominate across tasks.
In contrast, GPRF not only achieves the best overall performance but also consistently delivers the highest NDCG@10 across all datasets. 
%For example, on the TREC-COVID dataset, it improves over the vanilla BM25 by 30.1\%.
On the other hand, for the two dense retrievers, similar trends hold. 
GPRF not only surpasses VPRF but also outperforms strong GRF baselines such as HyDE and Lamer, with gains most evident on domain-shifted datasets like DBPedia and FiQA-2018, still achieving the highest overall scores in both NDCG@10 and R@100.
An additional advantage of GPRF lies in its domain-agnostic prompt design. 
While GRF methods like HyDE and Lamer require carefully crafted prompts tailored to different tasks or domains, GPRF employs a single unified prompt across all datasets. 
This not only simplifies deployment but also demonstrates the robustness and generalizability of our approach in cross-domain retrieval scenarios.

\begin{figure}[t]
    \centering
    \caption{Bucket-based analysis on MS dev. Queries are grouped into buckets based on their baseline BM25 performance, and the NDCG@10 improvement of three feedback-based methods, RM3, Lamer, and GPRF, is evaluated within each group. From left to right, the relevance of top-retrieved feedback documents in each group increases.}
    \includegraphics[width=0.95\columnwidth]{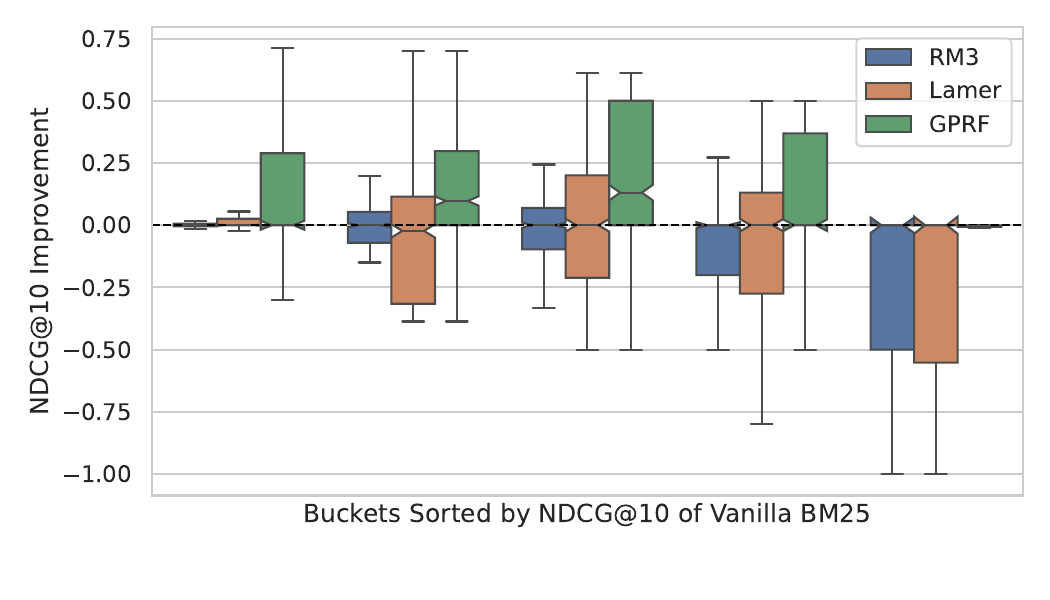}
    \label{fig:box}
\end{figure}

\begin{figure*}[htbp]
    \centering
    \caption{A case study shows that our training framework can effectively alleviate the noisy feedback problem.}
    \includegraphics[width=0.98\textwidth]{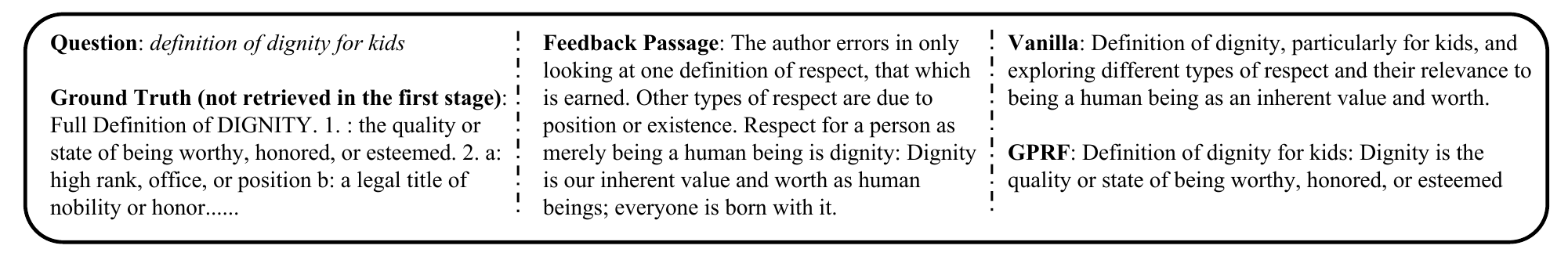}
    \label{fig:case}
    %\vspace{-6mm}
\end{figure*}

In summary, our GPRF framework demonstrates consistent and substantial improvements over classical PRF methods and recent GRF baselines across both in-domain and out-of-domain settings. 
In in-domain settings, GPRF leverages retrieval grounding to achieve significant gains in precision-oriented metrics (i.e., NDCG@10), while maintaining strong recall. 
In out-of-domain scenarios, it further shows superior robustness and adaptability, outperforming other GRF baselines without the need for task- or domain-specific prompt engineering. 
It not only highlights its strong capabilities and superior performance but also underscores its generalizability, demonstrating the ability to fully leverage the powerful natural language understanding and generation capacities of LLMs to achieve robust adaptation across different tasks and models, thereby effectively addressing the model assumption problem.
Besides, compared with other GRF methods, GPRF does not generate additional intermediate results or introduce extra modules, and thus maintains comparable efficiency.
Taken together, these results establish GPRF as a practical and effective solution for query rewriting, capable of delivering reliable performance across diverse retrieval settings.

\subsection{Ablation Study} 
To demonstrate the effectiveness of our utility-oriented training pipeline in tackling noisy feedback, we conduct ablation experiments by training the LLM with each stage individually, and the results are reported in Table~\ref{tab:abl}.
Compared to the vanilla setting without training, incorporating either SFT or RL alone yields substantial improvements, confirming the necessity of training for robust query reformulation: SFT provides a strong initialization, while RL contributes more significantly to performance gains.
Combining both within the full GPRF pipeline achieves the best overall performance, generally outperforming the single-stage and non-training variants.
By filtering low-utility generations through rejection sampling, grounding reformulations with SFT, and reinforcing utility-driven objectives via RL, GPRF reduces the negative impact of irrelevant or misleading feedback (we will further discuss later in \S\ref{subsec:case}), thereby alleviating the fragile relevance assumption. 
This allows the system to generate reformulations faithful to user intent.

\subsection{Relevance Assumption Analysis (RQ2)} \label{subsec:case}
To examine whether GPRF can effectively alleviate the relevance assumption, we conduct a bucket-based analysis on the MS dev dataset. 
Queries are grouped into buckets according to their baseline BM25 performance ordered from low to high, and we compare three feedback-based rewriting methods, RM3, Lamer, and GPRF, on each bucket, as shown in Figure~\ref{fig:box}. 
The results reveal a clear trend: while other methods exhibit limited or even negative gains on queries that already perform well (rightmost buckets), GPRF consistently yields substantial improvements, especially in the more challenging buckets where the baseline retriever performs poorly (left regions, indicating lower-quality feedback). 
In particular, the median NDCG@10 improvement of GPRF is significantly higher than RM3 and Lamer, indicating its stronger resilience to noisy or unreliable feedback.

A case study between the vanilla model without training and GPRF further (as shown in Figure~\ref{fig:case}) illustrates how GPRF alleviates the relevance assumption. 
For the given query "definition of dignity for kids", the truly relevant definition of "dignity" is not retrieved, and the top-ranked feedback document contains only partial or noisy signals.
It can be observed that the vanilla model is distracted by the feedback and produces a query expansion dominated by the notion of "respect", which drifts away from the canonical definition.
In contrast, GPRF can utilize the feedback context more effectively, filtering out spurious associations and grounding the reformulation in the core semantic meaning of "dignity".
%We observe that the vanilla model is distracted by the noisy feedback and produces a query expansion dominated by the noisy notion, which drifts away from the original user intent.
%In contrast, GPRF can utilize the feedback context more effectively, filtering out spurious associations and grounding the reformulation in the core concept.
This shows that our utility-oriented training pipeline enables the model to not only extract useful signals from noisy feedback documents but also to integrate them with its internal knowledge, producing a precise and faithful reformulation aligned with the user’s intent. 
It demonstrates how GPRF alleviates the fragile relevance assumption and ensures robustness in realistic retrieval scenarios where top-ranked documents may not be fully reliable.

To sum up, GPRF is capable of generating robust and semantically grounded reformulations even when the initial retrieval results contain substantial noise. 
In these realistic situations, traditional PRF and GRF methods tend to fail. 
By integrating retrieval grounding with our utility-oriented training pipeline, GPRF learns to selectively leverage relevant evidence while suppressing misleading information. 
Consequently, it effectively relaxes the dependence on the relevance assumption, maintaining stable gains across varying retrieval quality levels and demonstrating its robustness in real-world noisy retrieval environments.

\begin{figure}[t]
    \centering
    \caption{Cross-model experimental results on DL19 and DL20 with Llama. The results of providing different retrievers with various feedback are reported. It can be observed that using varying feedback consistently improves the performance of different retrievers.}
    \includegraphics[width=0.95\columnwidth]{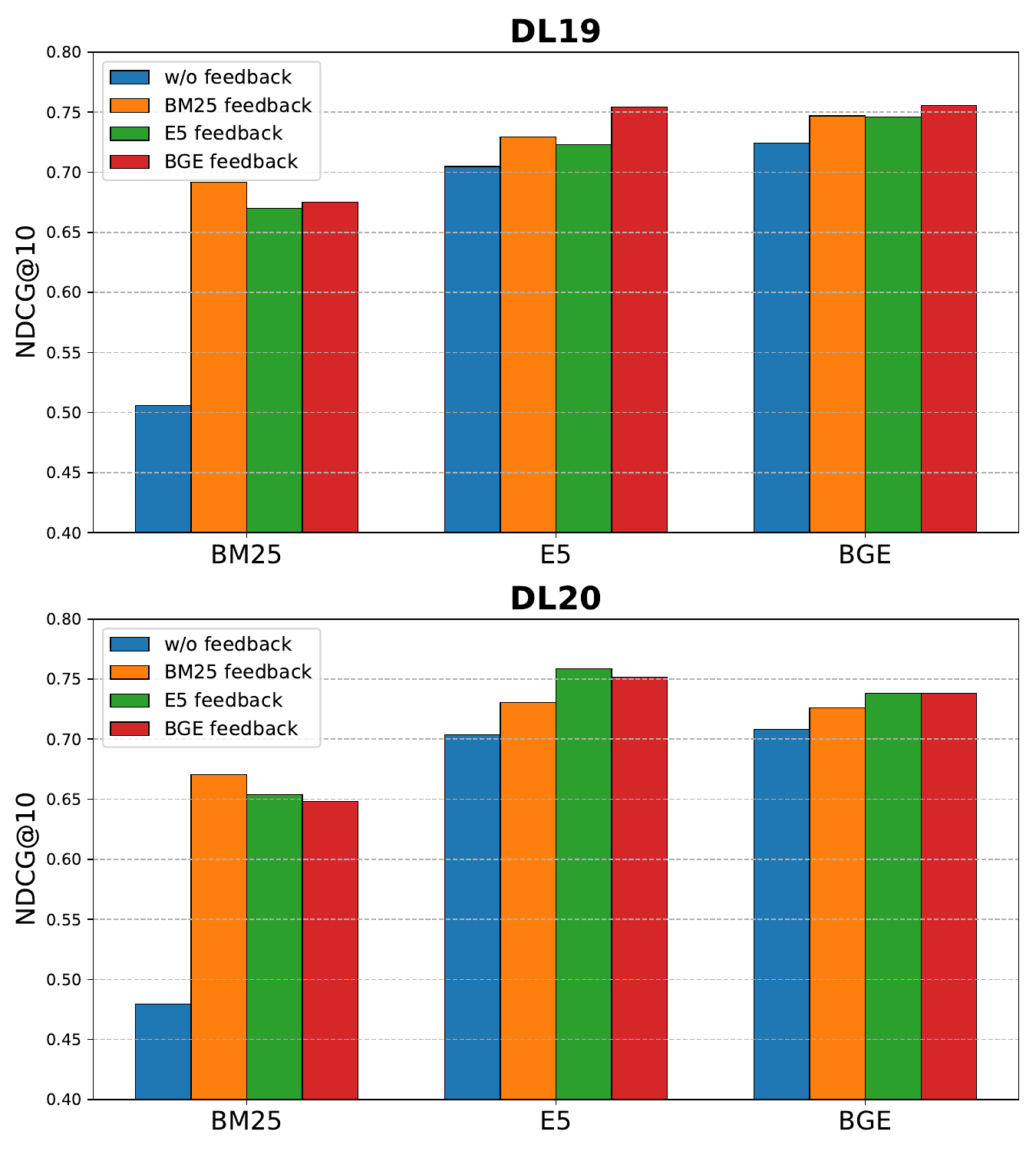}
    \label{fig:heat}
\end{figure}

\subsection{Cross-model Experiment (RQ3)}
On the other hand, to further verify that GPRF can effectively address the model assumption problem, we conduct cross-model experiments, where different retrievers are used to provide feedback documents and employed for the final retrieval.
As shown in Figure~\ref{fig:heat}, the cross-model results on DL19 and DL20 demonstrate that GPRF consistently achieves strong performance even when the retriever used for providing feedback differs from the one used for final retrieval, showing only minor or no performance drops compared to the in-model setting.
For instance, queries rewritten with BM25 feedback remain highly effective when evaluated with E5 or BGE, and using E5 as the feedback retriever still yields competitive results with BGE as the ultimate retriever.
Compared to the capability of a retriever itself, the impact of feedback retrievers is relatively insignificant.
Notably, even though BGE’s retrieval results or reward signals are never used during training, GPRF still achieves competitive performance when evaluated with BGE (also shown in Table~\ref{tab:ind} and Table~\ref{tab:ood}) or using it as the first-stage retriever.
All these indicate that GPRF-generated reformulations are not tied to the embedding space of any particular retriever, enabling robust transferability across heterogeneous retrieval models, and thereby validating the effectiveness in overcoming the model assumption and the generalizability of GPRF.

\section{Related Work}
\subsection{Ad-hoc Retrieval and Relevance Feedback}
Ad-hoc retrieval, the task of selecting documents according to their relevance to a given query, has been a central problem in information retrieval~\cite{guo2016deep,su2024wikiformer}. 
Traditional sparse retrieval methods, such as BM25~\cite{robertson2009probabilistic}, often rely on exact lexical overlap between queries and documents, with relevance scores determined by term frequency (TF) and inverse document frequency (IDF) statistics. 
While these approaches are computationally efficient and interpretable, they inherently lack deeper semantic understanding, often failing when queries and documents use different surface forms to express the same concept.
With the advent of pre-trained language models (PLMs) such as BERT and RoBERTa~\cite{devlin2019bert,liu2019roberta}, dense retrieval has emerged as a powerful alternative.
Dense retrievers~\cite{xiong2020approximate,izacard2021unsupervised,karpukhin2020dense,zhan2020repbert,fang2024scaling} map queries and documents into a shared semantic space, where relevance is measured by vector similarity. 
This paradigm enables retrieval beyond exact term matching, capturing semantic relations and paraphrases that sparse methods typically miss.
Despite their differences, both sparse and dense retrieval fundamentally operate within their respective vector spaces: sparse retrieval in high-dimensional lexical space, and dense retrieval in dense semantic space.
Their reliance on model-specific representations constrains traditional feedback methods such as PRF and VPRF~\cite{abdul2004umass,li2023pseudo}, which remain tightly coupled to a specific retrieval model  (i.e., the model assumption). 
In contrast, our proposed GPRF framework reformulates queries directly in natural language, providing a model-agnostic bridge that connects both sparse and dense retrieval, enabling more generalizable and transferable improvements across heterogeneous retrieval paradigms.

\subsection{Large Language Models (LLMs) for Query Rewriting}
Recent advances in large language models (LLMs)~\cite{achiam2023gpt,bai2023qwen,dubey2024llama} have opened new opportunities for information retrieval (IR) modules, including rewriter, retriever, reranker, and reader~\cite {zhu2023large}. 
In this paper, we mainly focus on the query rewriter module.
Representative work includes HyDE~\cite{gao2023precise}, which generates hypothetical documents as supplements for retrieval, and Query2Doc~\cite{wang2023query2doc}, which leverages few-shot prompting to produce answer-like passages as query expansions.
Similarly, chain-of-thought (CoT)~\cite{wei2022chain} prompting has been applied to query rewriting~\cite{jagerman2023query}, encouraging models to provide intermediate reasoning steps that lead to more interpretable reformulations.
These methods demonstrate that generation-based pseudo-documents can significantly improve retrieval coverage and recall.
Meanwhile, drawing from retrieval-augmented generation (RAG)~\cite{lewis2020retrieval,su2024dragin,su2025dynamic,su2025parametric}, other studies integrate retrieved documents into LLM prompting to mitigate generation bias and enhance factual accuracy. 
For instance, LameR~\cite{shen2023large} adopts a retrieve–answer–retrieve framework to guide LLM generation with pseudo-answers.
Although such methods can mitigate the hallucination problem of LLM outputs to some extent, their effectiveness is quite limited, especially when the retrieved results contain noise, as LLMs are vulnerable to noisy inputs~\cite{tu2025robust,wang2024astute}.
Building on this line of research, by reformulating queries with retrieval augmentation and leveraging a tailored training pipeline, our GPRF achieves greater robustness and generalizability than existing approaches while better relaxing the fragile relevance assumption.
\section{Conclusions}
In this paper, we revisit the limitations of classical query rewriting approaches, including PRF and VPRF, and recent LLM-based generative methods.
We identify two core challenges, the relevance assumption and the model assumption, that hinder their robustness and generalizability.
To address these issues, we propose Generalized Pseudo-Relevance Feedback (GPRF), unifying the strengths of retrieval-based grounding and LLM-driven generation, thus resolving the model assumption.
We also introduce a utility-oriented training pipeline to equip LLMs with the defense against noisy feedback and strengthen reformulation quality, thus alleviating the relevance assumption.
In future work, we plan to explore more effective and efficient training strategies and extend GPRF to multi-modal and interactive retrieval scenarios, further enhancing its applicability in real-world search systems.

\bibliographystyle{ACM-Reference-Format}
\bibliography{sample-base}
\end{document}